\documentstyle[aps,preprint,psfig]{revtex}
\newcommand{\beq}{\begin{equation}}
\newcommand{\eeq}{\end{equation}}
\newcommand{\beqa}{\begin{eqnarray}}
\newcommand{\eeqa}{\end{eqnarray}}
\newcommand{\beqan}{\begin{eqnarray*}}
\newcommand{\eeqan}{\end{eqnarray*}}
\tighten
\begin{document}

\vskip 1.0cm
\begin{center}
{\large \bf Kramers Equation Algorithm
 with Kogut-Susskind Fermions on Lattice}\\

\vskip 1.0cm 

{\sf \bf S. Basak \footnote{e-mail: basak@tnp.saha.ernet.in} and 
Asit K. De \footnote{e-mail: de@tnp.saha.ernet.in} }\\

\vskip 0.7cm 

{\sl Theory Group, Saha Institute of Nuclear Physics,\\
1/AF Salt Lake, Calcutta - 700 064, India}\\
\end{center}

\begin{abstract}
We compare the performance of the Kramers Equation 
Monte Carlo (KMC) Algorithm with that of the Hybrid Monte Carlo (HMC)
algorithm for numerical
simulations with dynamical {\em Kogut-Susskind fermions}. Using the lattice
Gross-Neveu model in 2 space-time dimensions, we calculate the
integrated autocorrelation time of different observables 
at a number of couplings in the
scaling region on $16^2$ and $32^2$ lattices while varying the parameters of
the algorithms for optimal performance.
In our investigation the performance of KMC is always significantly below 
than that of HMC for the observables used.
We also stress the importance of having a large number of configurations for
the accurate estimation of the integrated autocorrelation time.
\end{abstract}


\newpage

\noindent
{\bf Introduction.}
Numerical simulation is employed as a significant tool
in the nonperturbative investigation of lattice gauge theories. 
Most of the serious numerical simulations with dynamical lattice fermions
have so far been carried out with the so-called Hybrid Monte Carlo
(HMC) algorithm \cite{leap}. It has several virtues: it is an exact
algorithm, easily implementable and reasonably efficient
with a dynamical critical exponent $z=1$ in the large trajectory limit. 
However, with
growing demand of more realistic computations, especially in QCD, there is
a genuine need to search for better algorithms. There is also some  evidence 
\cite{karl,liu} of lack of
reversibility in the discretized classical equations of motion in the 
molecular dynamics part of the HMC. This will make the algorithm inexact. 
Since this will presumably be worse for larger lattice volumes, 
it is an unwelcome situation for future computations.

The main alternatives to HMC, available at the moment, are the local bosonic 
algorithm by Luescher \cite{luescher} and the Kramers Equation Monte
Carlo (KMC) \cite{karl,curci}, first introduced by Horowitz \cite{kram}. 
In this letter, we concentrate on the KMC algorithm which
is a generalized version of the HMC algorithm and is exact. 
The guiding trajectory of KMC includes a dissipative term. 
The optimally tuned KMC also maintains a dynamical critical exponent
$z=1$, but unlike HMC, for arbitrarily short trajectory length, at least for
free field theories. In practice, the
trajectory length is taken to be a single time-step and this presumably
would help  suppress
the effects of the possible lack of reversibility.

For this reason, the KMC certainly merits serious investigation and has 
received some
attention recently \cite{karl,curci,meyer} where KMC has been
applied to QCD with the gauge group SU(2) and to 1+1 dimensional Gross-Neveu
model. The general result of these investigations is that the KMC
has performed comparably with the HMC, although neither algorithm could be
called clearly more efficient. Apart from the preliminary report \cite{meyer}, 
the studies used {\em Wilson lattice fermions}. We feel that there is a
necessity for more investigation of KMC, especially with {\em Kogut-Susskind
fermions}. 

In the following we describe our investigation of the KMC algorithm
using {\em Kogut-Susskind}
or {\em staggered lattice fermions} on the 1+1 dimensional Gross-Neveu model.

\vskip 0.5cm
\noindent
{\bf The Gross-Neveu model.}
Since simulating QCD is computationally demanding, we have tested the
algorithms on a simple non-gauge model which shares some properties of QCD.   
The Gross-Neveu (GN) model \cite{gn} in 2 space-time dimensions 
is asymptotically free and
displays chiral symmetry breaking. Hopefully, our experience with the GN
model will serve us in good stead when confronted with the real problem, QCD.  

The action of the GN model on a 2-dimensional euclidean lattice can
be written down as:

\beq
S = \frac{1}{2g^2}\sum_x\sigma^2(x) + \sum_{x,y,f}
        \overline{\chi}_f(x)A_{xy}\chi_f(y)
\eeq
with,

\beqa
A_{xy} & = & \sigma(x)\delta_{xy} + \frac{1}{2}\sum_\mu\zeta_\mu(x)
             \{ \delta_{y,x+\mu} - \delta_{y,x-\mu} \}] \\
\zeta_1(x) & = & 1, ~~~~~~~~\zeta_\mu(x) = (-1)^{x_1+ \cdots + x_{\mu-1}},
\eeqa
where $\chi$ and $\overline\chi$ are the single-component Kogut-Susskind 
fermion fields; $\sigma$ is the scalar auxiliary field; $x$ and $y$ are
lattice sites and $f$ is a flavor index. If $n_f$
represents the number of fermion flavors on the lattice, 
the number of flavors in the continuum limit is $2n_f$. The lattice spacing
is taken to be unity. We note that the fermion matrix $A_{xy}$ is real and
independent of flavor.

The above action is the most naive transcription of the GN model on
a lattice. No improvement to reduce lattice artifacts has been used and the
$\sigma$-term in the fermion matrix $A_{xy}$ is taken to be single-site. 
However, this should be enough for our purpose.

\vskip 0.5cm
\noindent
{\bf Implementation of the algorithms.}  
We start by writing the classical Hamiltonian
\beq
H=\sum_{sites}\left[ \frac{1}{2g^2}\sigma^2 +\frac{1}{2}p^2 +\sum_f	
  \Phi^\dagger_f (A^\dagger A)^{-1} \Phi_f\right]
\eeq
which evolves the system in the fictitious time $\tau$ according to the
Hamilton's equations in the HMC algorithm. 
The $\Phi$ fields above are the usual pseudofermionic
fields, $p$'s are the momenta conjugate to $\sigma$ and are initially
obtained from respective gaussian distributions as suggested by the
quadratic terms in $\Phi$ and $p$ in $H$. The initial configuration of
$\sigma$ and $p$ are evolved in $\tau$ according to the Hamilton's equations
of motion in HMC while $\Phi$'s are kept fixed. The Hamilton's equations are
integrated in discrete steps of $\delta\tau$ using the leapfrog scheme for
$N_{md}$ steps to complete a molecular dynamics trajectory at the end of which
an accept/reject step is performed. This is well known and the details of
the HMC algorithm can be found in many excellent reviews and articles.

The KMC involves the modification where
a term proportional to momenta $p$ is added to the momenta itself in each
trajectory,

\beq
\pi_i = e^{-\gamma\delta\tau}p_i + \sqrt{1 - e^{-2\gamma\delta\tau}}\eta_i
\label{krmom}
\eeq

\noindent
where $\eta_i$ is a Gaussian noise with zero mean and unit variance
and $\gamma$ is a tunable parameter. 
The subscript $i$ refers to the number of trajectory. 
The
leap-frog integration is performed over a single molecular dynamics
step $(N_{md}=1)$:
\beqan
\pi_i(\delta\tau/2)&=& \pi_i(0) + \frac{\delta\tau}{2} 
F\left(\sigma_i(0)\right), \\  
\sigma_{i+1}(\delta\tau) &=& \sigma_i(0) + \delta\tau \pi_i(\delta\tau), \\
\pi_{i+1}(\delta\tau) &=& \pi_i(\delta\tau/2) + \frac{\delta\tau}{2} 
                    F\left(\sigma_{i+1}(\delta\tau)\right),
\eeqan
with the force at the site $x$
\beq
F_x(\sigma_i)= -\frac{1}{g^2}\sigma_i(x) +\sum_f\left[
\Omega_f^\dagger(x)\xi_f(x)+\xi^\dagger_f(x)\Omega_f(x)\right],
\eeq
where $\Omega$ and $\xi$ are given by, $(A^\dagger A) \xi_f = \Phi_f,\;\;
\Omega_f = A\xi_f$.

After this a Metropolis accept/reject test with acceptance probability
$P(\sigma_i,\pi_i \rightarrow \sigma_{i+1},\pi_{i+1})
= min\{1,e^{H(\sigma_i,\pi_i)-H(\sigma_{i+1},\pi_{i+1})}\}$
is performed. If the new configuration $(\sigma_{i+1},\pi_{i+1})$ is
accepted, we start again from Eq.~\ref{krmom} with $p_{i+1} = \pi_{i+1}$,
and if rejected, we set $\sigma_{i+1}=\sigma_i,\;\;p_{i+1}=-\pi_i$
and try again starting from Eq.~\ref{krmom}. It is important to negate the
momenta in case of rejection to guarantee exactness of the algorithm.
Momenta $p$ and the 
pseudofermionic fields $\Phi$ are refreshed after every $k$-trajectories.

Here $\gamma$ and $k$ are tunable parameters. In the limit 
$\gamma=\infty$ and $k=1$ KMC reduces to HMC algorithm with $N_{md}=1$.

\vskip 0.25cm
\noindent
{\bf Autocorrelation time.} A good measure of how correlated are the
configurations generated by HMC and KMC, for a particular observable, 
is the integrated autocorrelation time for that observable. For
either algorithm, we have measured it for $\sigma$ and the fermion propagator.

Using the {\it window} method suggested by Sokal \cite{acl}, an estimator 
for the integrated autocorrelation time $\tau_{int}$ of some observable 
$\theta$ can be expressed as

\beq
\tau_{int}(T_{cut}) = \frac{1}{2} + \sum_{t=1}^{T_{cut}} \frac{C_\theta(t)}
                    {C_\theta(0)}.  \label{tint} 
\eeq

\noindent
where $C_\theta(t)$, the unnormalized autocorrelation function, can be
estimated by the following expression:

\beq
C_\theta(t) = \frac{1}{n-|t|} \sum_{i=1}^{n-|t|} (\theta_i - \overline{\theta})
               (\theta_{i+t} - \overline{\theta}).
\eeq

\noindent
$\theta_i$ is the average value of $\theta$ for the i-th configuration 
and $\overline{\theta}$ is obtained by averaging over the $\theta_i$'s.
$n$ is the total number of configurations, a large number, but obviously
finite.
  
In the above, the factor of $1/2$ is purely a convention and 
$T_{cut}$ is a cut-off for the sum, to be chosen judiciously so that one
strikes a balance between noise (in case $T_{cut}$ is too large) and bias (in
case $T_{cut}$ is too small) in the estimation of $\tau_{int}$. In short, the
recommendation of ref.~\cite{acl} is to take $T_{cut} \geq c \tau_{int}$,
where c is a number between 4 and 10 depending on the system and
$n\gg\tau_{int}$.

If one plots $\tau_{int}$ as a function of $T_{cut}$ for a given
observable, one can read off the value of $\tau_{int}$ from
a plateau region where $\tau_{int}$ is independent
of $T_{cut}$. Such plateau regions are indeed obtained and will be discussed
when results are presented in the following.

\vskip 0.5cm
\noindent
{\bf Performance tests.}
The numerical simulations were carried out on $16^2$ and $32^2$ lattices
using $n_f=4$. For HMC, we maintained $N_{md}\delta\tau\sim 1$ and the ansatz
for the input vector for the matrix inversion at each time-step within the
molecular dynamics trajectory was simply taken to be the result-vector of the
matrix-inversion of the previous
time-step. For both algorithms, the parameters of the guiding Hamiltonian
were not tuned for optimal performance and the conjugate gradient inverter was
used for fermion matrix inversions without any preconditioning.
Preconditioning does not seem to work anyway for Kogut-Susskind fermions.
The parameters of the algorithms were always tuned so as to keep the
acceptance rate above 90\%.

We tried both cold and hot starts, but for the production runs we always
used cold starts for both algorithms. At each set of parameters we
typically used 1000 iterations in HMC and 16000 iterations in KMC for
equilibration. As we will see, these numbers are substantially larger than
$20 \tau_{int}$, suggested by \cite{acl}, for the observables of interest.

We determined the scaling region by plotting $\langle\sigma\rangle$ as
a function of $1/g^2$. The scaling window 
is between $g=0.55 - 0.45$ for $16^2$ lattice and gets slightly  beyond 
0.45 for $32^2$. For the investigation of autocorrelation times, we
collected data at various values of the coupling $g$ always remaining within
the scaling window.

For HMC, we varied $\delta\tau$ from 0.033 to 0.05 always maintaining
$N_{md}\delta\tau\sim 1$. For each coupling we tried to find the best value
for the ratio $r=a/2\tau_{int}$ where $a$ is the acceptance rate. The bigger
the ratio, the better is the performance. On $16^2$ lattice 
at $g=0.5$ this is obtained at
$\delta\tau=0.04$ for the observable $\langle\sigma\rangle$ and roughly also
for the fundamental fermion propagator. In Fig.1a we have plotted
$\tau_{int}$ versus $T_{cut}$ for each of these observables for $16^2$
lattice. The lower curve
is for the fermion propagator, and to be specific, we have presented the
autocorrelation data for zero-spatial-momentum propagator at time separation
of 3. For the subsequent plots of fermion propagator autocorrelation, we
will use this particular propagator. For a given $T_{cut}$, we have measured
$\tau_{int}$ from each of 15 segments, each segment-size being 1500
configurations for
$\langle\sigma\rangle$ and 1000 for the fermion propagator. The errors on the
$\tau_{int}$ data represents the statistical error from the 15 values for
each $T_{cut}$. In subsequent plots the same procedure will be followed,
although the segment-size and the number of segments will be changed
appropriately. From Fig.1a, one can easily find a plateau region from which
$\tau_{int}$ can be read off. One sees also that the fermion propagator
has significantly smaller integrated autocorrelation time.

For a reasonably accurate determination of $\tau_{int}$ we found it
absolutely essential to have large segment sizes ($\sim 1000\tau_{int}$).
With smaller segment sizes we have found it difficult to find the plateau
because of excessive noise. As a general observation, we have noticed that
the value of $\tau_{int}$, if at all one can read it off, is always lower
when the segment-size is small. In HMC, for both $16^2$ and $32^2$ lattices,
we have actually determined $\tau_{int}$ at
least for $\langle\sigma\rangle$ for a host of other values of
$g=0.52,\;0.48,\;0.46$ with segment-size of 100 configurations and have
rough estimation of autocorrelations at those couplings too. For KMC, the
ill-effects of having small segment sizes were more dramatic.

For KMC, we varied $\delta \tau$ from 0.033 to 0.05 on $16^2$ and $32^2$
lattices and also varied the parameters $\gamma$ and $k$. Within our range
of the parameters the near-optimal choice for the ratio $r$ at $g=0.5$ was
$\delta\tau=0.04,\;\;\gamma=1.5,\;\;k=4$  on $16^2$ lattice and in 
Fig.1b we present the corresponding autocorrelation time data (later, we
found that by increasing $k$ to 8, it was possible to enhance the
performance of KMC somewhat, and thus our choice of parameters is not quite
optimal). 
Large statistics was required to produce the figure. The segment-size was
72,000 configurations for fermion propagators and 100,000 for
$\langle\sigma\rangle$, with 5 such segments for each observable. Once again,
nice plateau regions were found for $\tau_{int}$. The propagators are again
less correlated, but the values of $\tau_{int}$ for either observable in the
plateau region are orders of magnitude higher than those obtained in Fig.1a
for HMC. The acceptance rates are usually always higher for KMC, but the
huge discrepancy in the autocorrelations, even for observables like the
propagators, makes the HMC significantly better in effective performance, in
spite of HMC's higher overhead per trajectory (see Table 1 below). 

We made a qualitative estimate of the higher overhead of HMC over KMC per
trajectory for all our runs. Ratio of total number of conjugate gradient
iterations for one full trajectory of HMC to that for one trajectory of KMC
(including Hamiltonian calculations in each case) gives an accurate estimate of
the overhead. For the runs in Figs.1a and b, this overhead is 14.0 with
10\% uncertainty. For all our runs at different couplings, lattice sizes and
algorithm parameters, the HMC overhead was between 10 and 15. 

To illustrate the point of having large enough segment sizes, we plot in
Fig.2 again $\tau_{int}$ versus $T_{cut}$ for $\langle\sigma\rangle$
on $16^2$ lattice at $g=0.5$ using
the same set of parameters, but this time using 6 segments each of size 1000
configurations only. Please note the $\tau_{int}$-scale of Fig.2 as compared
to that of Fig.1b. There is only a rough hint of a plateau in Fig.2 and the
data is very noisy. Also, the actual value of $\tau_{int}$ at the
anticipated plateau region of Fig.2 is about half of that in Fig.1b.

In KMC we have investigated integrated autocorrelation 
times at various other couplings,
{\it viz.}, $g=0.52,\;0.48,\;0.46$ with $\gamma=1.5$, $\delta\tau=0.04$ and
$k=4$ on both $16^2$ and $32^2$ lattices, however, 
with smaller statistics as in Fig.2
and could estimate $\tau_{int}$ for the observables only very roughly. Still
in each case the autocorrelation time was found to be orders of magnitude
higher than the corresponding values with HMC, as shown in Table 2 presented
later. 

On $16^2$ lattice, with $k=4$ and $\delta\tau=0.04$ 
we have also varied the parameter $\gamma$ 
from 0.5 to 2.0 in intervals of 0.5  for a number couplings $g$ from 0.52 to
0.46. The autocorrelation time determinations could only be done with
relatively small statistics. The optimal $\gamma$ was found in each case to
be qualitatively consistent with the free theory estimate of $1/m_s$ where
$m_s$ is the smallest mass of the theory. As a result the optimal $\gamma$
shifted towards higher values as the coupling $g$ was lowered with all other
parameters kept fixed.

In addition to $k=4$, we have performed similar autocorrelation
investigations at $k=8$ on $16^2$ lattice using $\gamma=1.5$,
$\delta\tau=0.04$ at couplings $g=0.50,\;0.46$. By increasing the value of
$k$, there was an indication in our data of an enhancement of the
ratio $r$ at the lower coupling. Still it was too little to
boost it up on par with HMC.

In Table 1 we present the values of $\tau_{int}$ and the ratio $r$ 
for $\langle\sigma\rangle$ and the fermion propagator at $g=0.5$ on $16^2$
lattice in
both HMC and KMC for the data shown in Fig.1a and Fig.1b. Put together with
higher overhead for HMC (dicussed earlier), table 1 then indicates that HMC
is about 9.5 and 7.5 times more efficient than KMC respectively for $\langle
\phi \rangle$ and for fermion propagator.
Table 2 collects
some results from small statistics runs on both $16^2$ and $32^2$ lattices
at a few other couplings in the scaling region. The data at $g=0.46$ on
$16^2$ lattice has higher statistics than the rest in Table 2, as evident
from smaller errors. The KMC data on $32^2$ lattice are the most inaccurate
and the values of $\tau_{int}$ are likely to be significantly below their
actual values.   

\vskip 0.25cm
\noindent
{\bf Conclusions.}
In our investigation of relative performances of HMC and KMC algorithms with
dynamical {\em Kogut-Susskind fermions} on a simple 2-dimensional
Gross-Neveu model, we found that the HMC was significantly 
more efficient in producing
uncorrelated configurations than the KMC. When
other similar investigations \cite{karl,curci} with {\em Wilson fermions}
did not find a significant difference of performance between the two
algorithms, our results indicate, taking into account the higher overhead
per trajectory of the HMC which in our case was always between 10 and 15, 
a difference of about {\em an order of magnitude} 
for all
combinations of parameters and couplings  
we tried. How much of our conclusion would carry over to QCD with
Kogut-Susskind fermions is an open question. 

Our choice of parameters for KMC might not have been optimal in
all cases, but it becomes apparent that KMC requires a lot of tuning and our
investigation shows that we were unable to find a set of parameters which
could make KMC even remotely competitive.   

We emphasize again the importance of having a large number of configurations
for the estimation of $\tau_{int}$. Without it, the data is not just noisy,
the roughly estimated $\tau_{int}$ is well below the actual value.


\vskip 2cm
\begin{center}
{\large FIGURE CAPTIONS}
\end{center}

\noindent
{\bf Fig. 1.} Integrated autocorrelation times for $\langle\sigma\rangle$
and fermion propagator (zero spatial momentum with time-separation 3) 
at $g=0.5$ on $16^2$ lattice for (a) HMC with $\delta\tau=0.04$,
(b) KMC with $\delta\tau=0.04, \; \gamma=1.5, \; k=4$.

\vskip 1.5cm

\noindent
{\bf Fig. 2.} Low statistics data on $\tau_{int}$ for $\langle\sigma\rangle$
in KMC at $g=0.5$ on $16^2$ lattice with the same set of parameters as in
Fig. 1b.


\begin{center}
{\large TABLE CAPTIONS}
\end{center}

\noindent
{\bf Table 1.} $\tau_{int}$ and $r$ for $g=0.50$ obtained on
$16^2$ lattice (HMC: $\delta\tau =0.04$; 
KMC: $\delta\tau =0.04$, $\gamma=1.5$, $k=4$).

\vskip 1.5cm

\noindent
{\bf Table 2.} $\tau_{int}$ and $r$ for various $g$ obtained on
$16^2$ and $32^2$ lattice for $\langle\sigma\rangle$ (HMC: $\delta\tau =0.04$; 
KMC: $\delta\tau =0.04$, $\gamma=1.5$, $k=4$).

\newpage

\begin{center}

{\large TABLES}

\vskip 1.5cm

\begin{tabular}{|c|c|c|c|c|} \hline 
Algorithm & 
\multicolumn{2}{c|}{$\langle\sigma\rangle$} & 
\multicolumn{2}{c|}{$propagator$} \\ \hline
& {\small $\tau_{int}$} & {\small $r$} & {\small $\tau_{int}$} & {\small $r$}
   \\ \cline{2-5}
& & & & \\
HMC & 1.72(28) & 29(5) & 0.83(11) & 63(8) \\
& & & & \\
KMC  & 228(27) & 0.22(2) & 83(6) & 0.60(5) \\
& & & & \\ \hline
\end{tabular}

\vskip 0.7cm
TABLE 1.

\vskip 2.0cm

\begin{tabular}{|c|c|c|c|c|c|c|c|} \hline
\multicolumn{2}{|c|}{} &
\multicolumn{2}{c|}{$g=0.52$} & 
\multicolumn{2}{c|}{$g=0.48$} &
\multicolumn{2}{c|}{$g=0.46$} \\ \cline{3-8}
\multicolumn{2}{|c|}{} 
 & {\small $\tau_{int}$} & {\small $r$}
 & {\small $\tau_{int}$} & {\small $r$}
 & {\small $\tau_{int}$} & {\small $r$} \\ \hline
 & & & & & & & \\
 & HMC & 1.22(48) & 45(18) & 1.30(45) & 40(13) & 1.05(20) & 48(9) \\
$16^2$ & & & & & & &\\
 & KMC & 128(64) & 0.50(25) & 108(26) & 0.47(11) & 138(23) & 0.36(6) \\ 
 & & & & & & & \\ \hline
 & & & & & & & \\
 & HMC & 1.47(38) & 33(8) & 1.47(50) & 33(11) & 1.56(31) & 30(6) \\
$32^2$ & & & & & & &\\
 & KMC & 124(97) & 0.84(66) & 142(107) & 0.63(47) & -- & -- \\ 
 & & & & & & & \\ \hline
\end{tabular}

\vskip 0.7cm
TABLE 2.
\end{center}

\newpage
\pagestyle{empty}

\begin{figure}
\parbox{7.5cm}{\psfig{figure=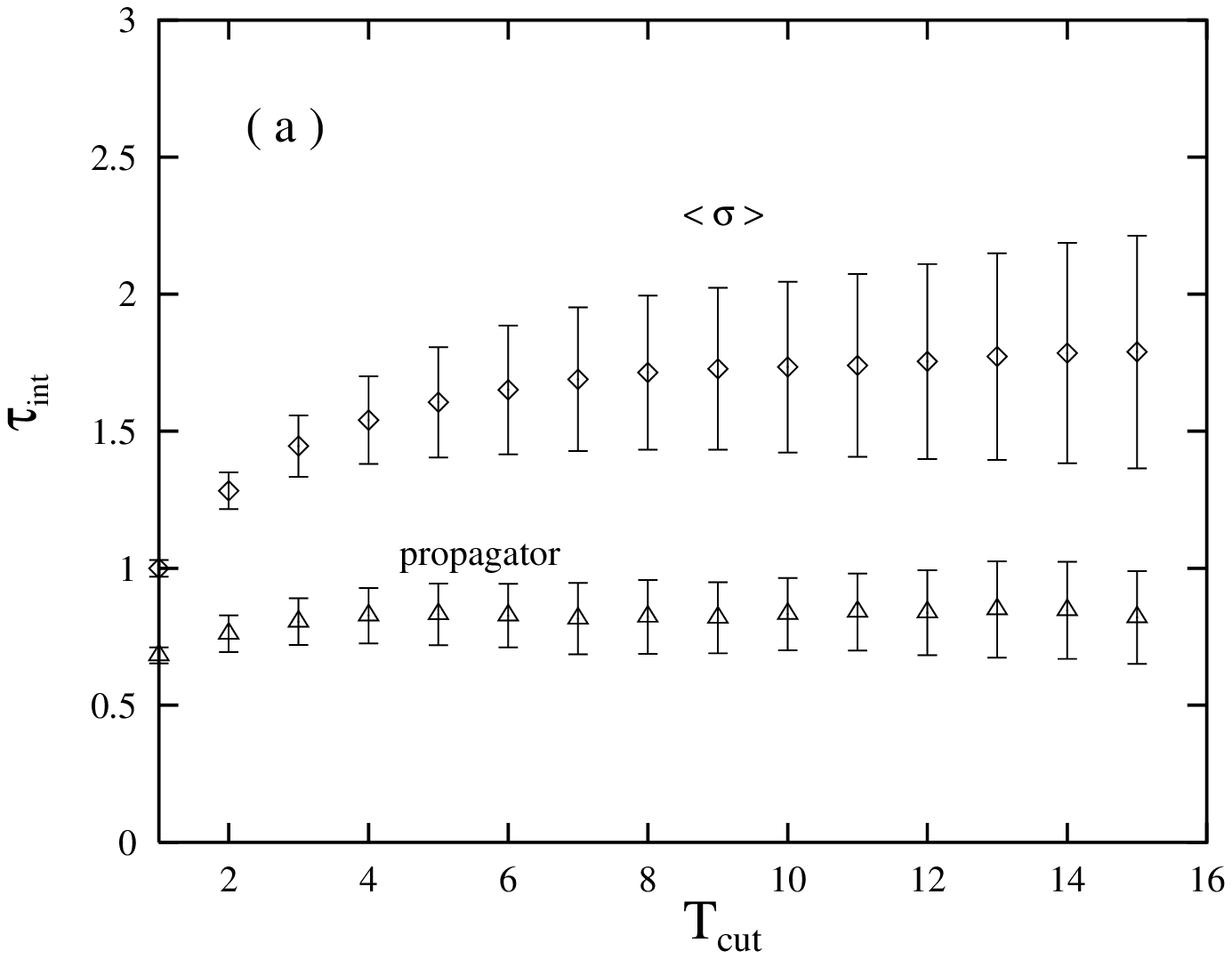,width=7.5cm,height=6.5cm}} \ \
\parbox{7.5cm}{\psfig{figure=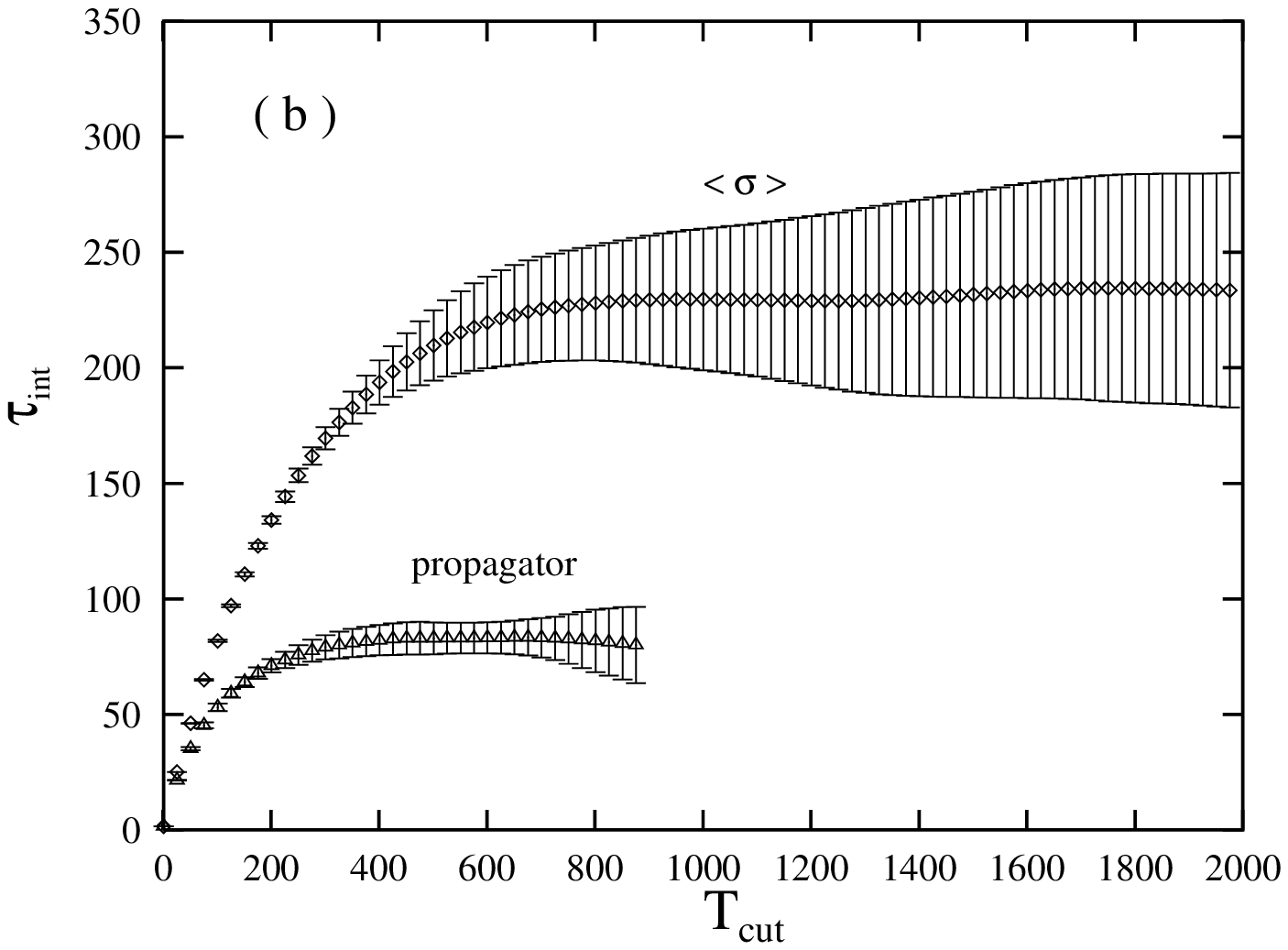,width=7.5cm,height=6.65cm}}

\vskip 0.7cm
\caption{}
\end{figure}

\vskip 1.0cm

\begin{figure}
\centerline{\psfig{figure=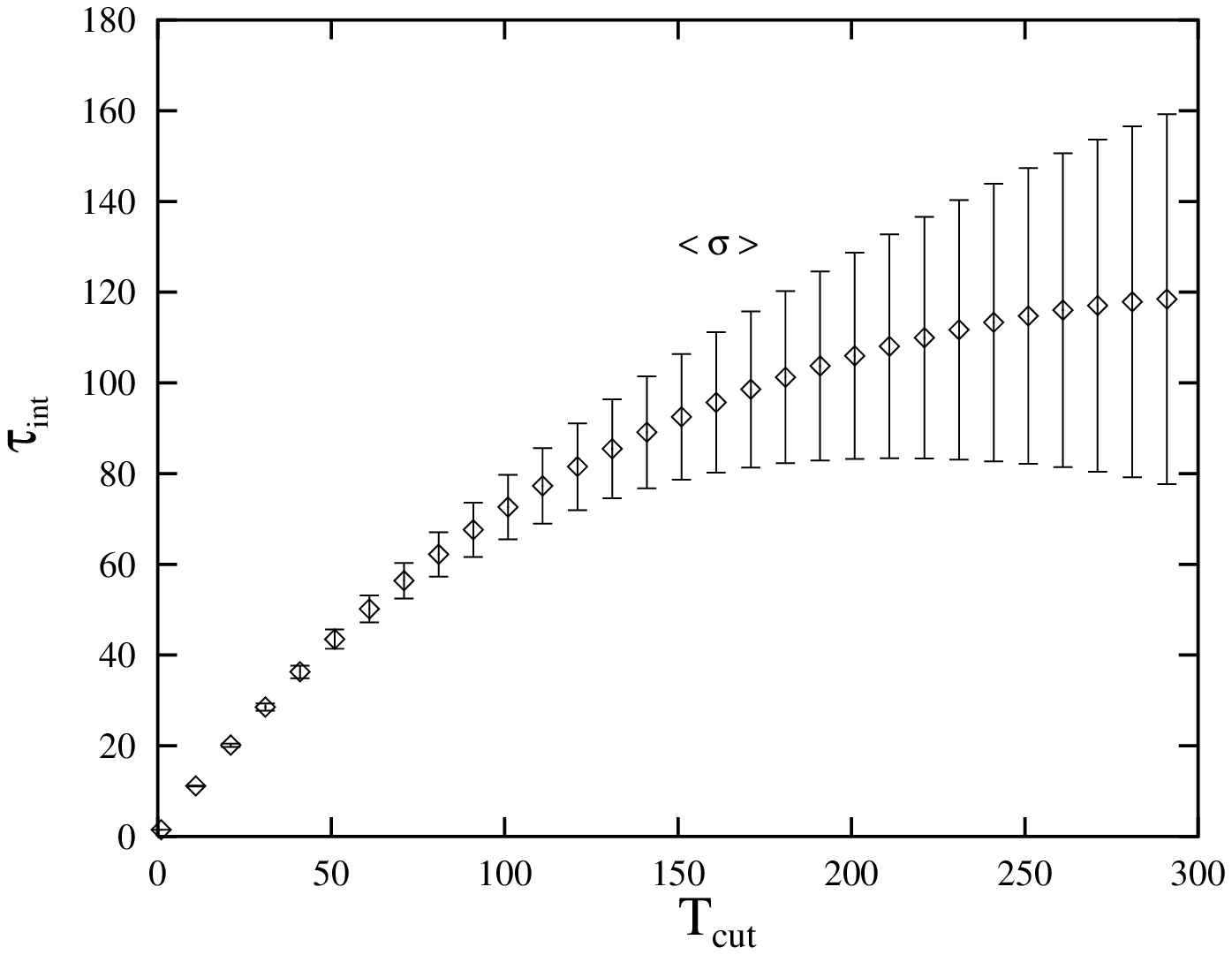,width=9.5cm,height=8.0cm}}

\vskip 0.7cm
\caption{}
\end{figure}

\end{document}